**Transparent AI: Developing an Explainable Interface for Predicting Postoperative Complications**

Yuanfang Ren, PhD[a,b], Chirayu Tripathi, BS[a], Ziyuan Guan, MS[a,b], Ruilin Zhu, MS[a,b], Victoria Hough[a], Yingbo Ma, PhD[a,b], Zhenhong Hu, PhD[a,b], Jeremy Balch, MD[a,c], Tyler J. Loftus, MD, PhD[a,c], Parisa Rashidi, PhD[a,d], Benjamin Shickel, PhD[a,b], Tezcan Ozrazgat-Baslanti, PhD[a,b*], Azra Bihorac, MD, MS[a,b*]

*These senior authors have contributed equally

[a] University of Florida Intelligent Clinical Care Center, Gainesville, FL.

[b] Department of Medicine, Division of Nephrology, Hypertension, and Renal Transplantation, University of Florida, Gainesville, FL.

[c] Department of Surgery, University of Florida, Gainesville, FL.

[d] Department of Biomedical Engineering, University of Florida, Gainesville, FL.

Corresponding author: Azra Bihorac, MD, MS, University of Florida Intelligent Clinical Care Center, Division of Nephrology, Hypertension, and Renal Transplantation, Department of Medicine, University of Florida, PO Box 100224, Gainesville, FL 32610-0224. Telephone: (352) 294-8580; Fax: (352) 392-5465; Email: abihorac@ufl.edu

**Abstract**

Given the sheer volume of surgical procedures and the significant rate of postoperative fatalities, assessing and managing surgical complications has become a critical public health concern. Existing artificial intelligence (AI) tools for risk surveillance and diagnosis often lack adequate interpretability, fairness, and reproducibility. To address this, we proposed an Explainable AI (XAI) framework designed to answer five critical questions — "why, why not, how, what if, and what else" —with the goal of enhancing the explainability and transparency of AI models. We incorporated various techniques such as Local Interpretable Model-agnostic Explanations (LIME), SHapley Additive exPlanations (SHAP), counterfactual explanations, model cards, an interactive feature manipulation interface, and the identification of similar patients to address these questions. We showcased an XAI interface prototype that adheres to this framework for predicting major postoperative complications. This initial implementation has provided valuable insights into the vast explanatory potential of our XAI framework and represents an initial step towards its clinical adoption.

**Introduction**

Among the 187.2 to 281.2 million surgeries conducted globally per year, approximately 7 million patients develop major complications, resulting in 1 million deaths during or immediately following surgery.[1] Given the sheer volume of surgical procedures and the significant rate of postoperative fatalities, assessing and managing surgical complications has become a critical public health concern. Additionally, there exists an unmet need for an objective method to identify and effectively manage patients who are at the highest risk of post-surgical complications.[2]

Preoperative assessment of surgical risk requires integration and interpretation of the large amount of clinical information scattered throughout the healthcare system, but their magnitude and complexity often overwhelms physicians' ability to comprehend, retain, and organize the information in an optimal and timely way. The widespread adoption of electronic health records (EHR) leading to digital data has enabled the proliferation of artificial intelligence (AI) tools for risk surveillance and diagnosis.[3-8] However, progress in the field remains halted by models with insufficient interpretability, fairness, and reproducibility that are difficult to implement and share across institutions.

The lack of explainability of model processes led to AI models (i.e., deep neural networks) being described as black-box models.[9] AI implementation in healthcare poses a challenge as explainability is the fundamental principle of evidence-based medicine.[10] For instance, clinicians may use AI models to supplement their clinical judgment when making a decision, but the lack of explainability and transparency is crucial in the high-risk medical environment where the potential of harm is vast.[11, 12] If the AI tool provides a confusing and untraceable decision, clinicians may lose trust in the model, abandon it, and forego a valuable supplemental tool that could benefit their patients.[9, 13] Thus, there remains a need for sequential and explainable models if AI systems are to be successfully implemented into clinical decision-making.

Explainable AI (XAI) represents a paradigm shift in machine learning (ML), offering models that not only predict but also provide insights into the underlying reasons for their predictions, characterizing real-world phenomena with greater clarity.[13] Some risk prediction models incorporated theory-based evaluation frameworks and the application of psychological and mental models to allow users to establish a mental model of the model's sequential decision-making process.[14, 15] In an acute kidney injury risk prediction model, the principal factors influencing its predictions, along with an associated uncertainty level were listed, thereby enhancing user comprehension.[16] Other approaches for explainability, such as SHapley Additive exPlanations (SHAP) and integrated gradients, have also been widely used in the recent clinical AI models to demonstrate feature importance to users and enhance the model's explainability.[6, 8, 15, 17, 18] While these instances represent just a fraction of the advancements in the field, further details and examples are well-documented in various review studies.[19, 20] Despite these strides in integrating explainability into AI risk prediction models, a more profound and comprehensive exploration of explainability within AI systems remains an imperative area for future research, to fully realize the potential of these technologies in practical applications.

The objective of this study is to propose the framework of an XAI model providing comprehensive knowledge of explainability, and design an XAI interface prototype for predicting major postoperative complications.

**Methods**

*XAI framework*

We have developed an XAI framework, specifically designed to unveil insights into decisions and actions derived from AI prediction model outputs. This framework methodically addresses five fundamental questions: "why", "why not", "how", "what if", and "what else" (Figure 1).

***"Why" question:*** Understanding "Why?" an AI model made a particular decision is fundamental in making the underlying inference process transparent and understandable. Such transparency is key to building trust and acceptance among users and stakeholders. In our XAI framework, the "why" question was answered by providing feature importance, which quantifies the impact of each input variable on the model's prediction. To achieve this, we employed techniques including Local Interpretable Model-agnostic Explanations (LIME)[21] and SHapley Additive exPlanations (SHAP)[22].

***"Why not" question:*** Understanding the rationale behind "Why not?" an AI model did not make a different outcome is equally imperative as comprehending the "Why" in the context of model decision-making processes. This is particularly crucial in scenarios where absence of a particular outcome needs explanation. In our XAI framework, the "Why not" question was answered by providing counterfactual explanations[23], which provide answers about what changes should be made to alter the model's decision.

***"How" question:*** The "How?" question pertains to understanding the operational and methodological aspects of an AI model, providing clarity on its development, performance, and use cases. Clear clarification about how a model functions enhances the model's transparency and builds trust among users. Our XAI framework addressed this problem by providing the model cards. Model cards are short documents accompanying AI models to disclose the context in which models are intended to be used, details of the evaluation procedures, and other relevant information. We augmented these model cards with additional layers of explainability, which includes presenting the global feature importance measuring the overall impact of each feature on the model's predictions.

***"What if" question:*** The "What if?" question involves exploring how changes in the feature values might affect the predictions of AI models. Such interactive explainability enables users to experiment and understand the model's behavior under various hypothetical scenarios. Our XAI framework answered this "What if" question through an interactive user interface which allows users to interactively manipulate feature values and runs inference on perturbed data points.

***"What else" question:*** The "What else?" question seeks to understand the model behavior on instances that are similar to the one being analyzed, thereby providing a broader understanding of the consistency of the model's decision-making process. In our XAI framework, the "What else" question was answered by providing model's predictions on similar instances. We generated similar instances by matching the principal features.

*XAI Interface for major postoperative complications*

Leveraging the XAI framework, we have developed a prototype interface for MySurgeryRisk, our prospectively validated surgical risk prediction model.[4] This model harnesses a random forest algorithm and integrates 135 input features from EHR data to generate patient-level probabilistic scores for eight major postoperative complications (acute kidney injury, sepsis, venous thromboembolism, prolonged intensive care unit admission, prolonged mechanical ventilation (MV), wound, neurologic, and cardiovascular complications), and 30 day and 90 day mortality after surgery. In this study, we refined MySurgeryRisk model by retraining it with a subset of features. We excluded highly correlated features, predominantly those derived from laboratory variables. That is, for each preoperative laboratory variable, we selected the most abnormal measurement (i.e., maximum or minimum) instead of utilizing all derived features (i.e., minimum, maximum, count, variance, and mean). The final feature list was detailed in eTable 1. The rationale behind these modifications was to enhance the model's capacity in answering "why not" and "what if" questions.

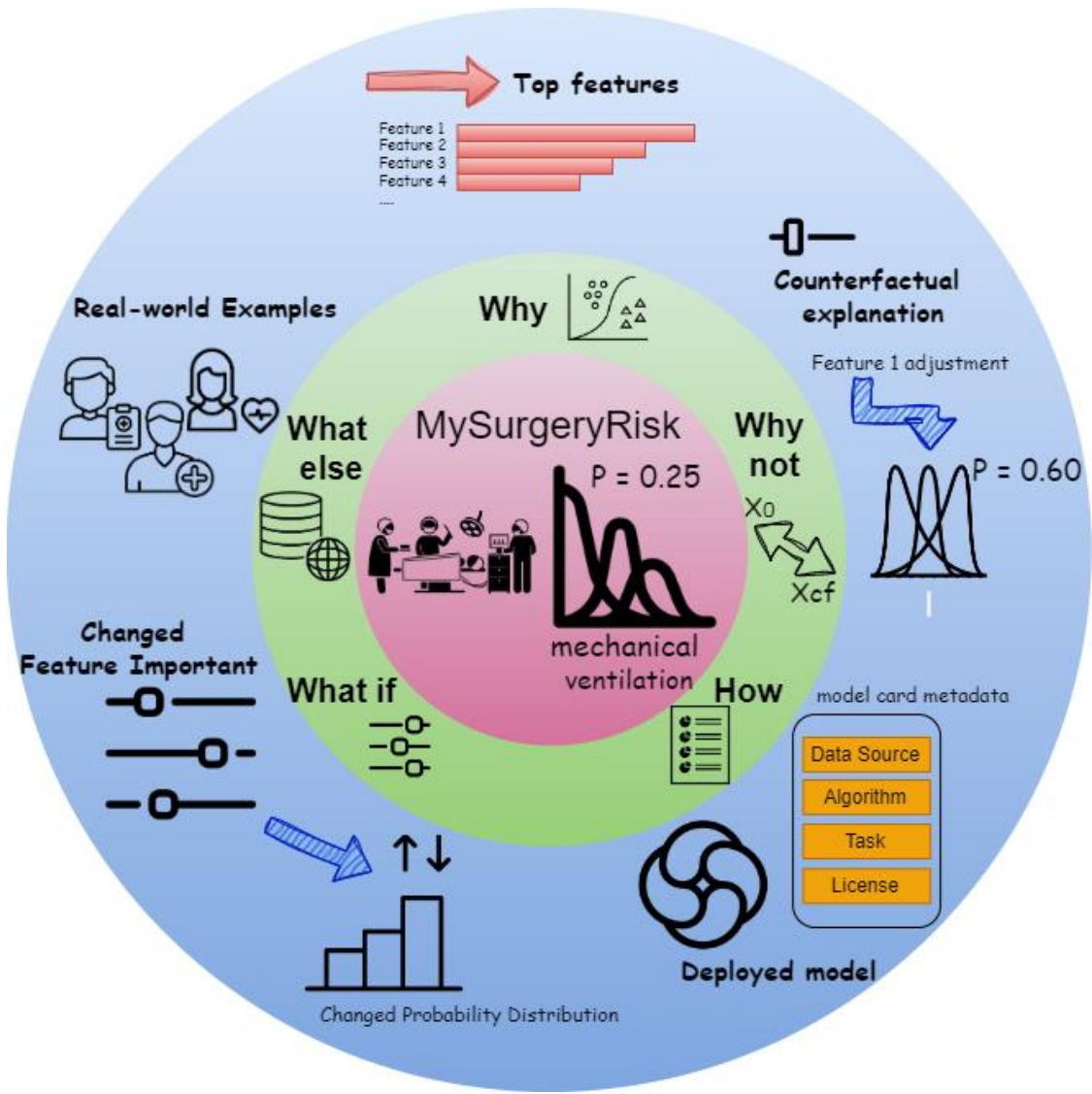

**Figure 1. Explainable Artificial Intelligence (XAI) framework.**

## Results

*Local feature importance explaining why a specific prediction was made*

We reported results on the explainability of predictions specifically for prolonged MV. Figure 2 and eFigure 1 present top 10 important features generated by LIME approach, illustrating their contributions to the predictions for two example patients: one predicted to have high probability (74%) of requiring prolonged MV, and the other with a low probability (12%).

The influence of these features on the predictions is indicated by their magnitude and direction. For example, for the patient at high risk (Figure 2), the scheduled primary procedure code 31600 that corresponds to under incision procedures on the trachea and bronchi is associated with an approximately 13% increased chance of requiring prolonged postoperative MV. Additionally, the presence of a comorbidity such as weight loss raises the probability by 4%, while a non-transfer hospital admission reduces it by 2%. For the patient with low risk (eFigure 1), several factors contribute to reducing their risk. For example, patient did not have comorbidities of weight loss or cerebrovascular disease, was not transferred from another hospital, and had normal preoperative calcium levels. However, some neighborhood characteristics slightly increased the risk.

We explored another approach SHAP to provide feature importance (eFigures 2 and 3). We noted significant differences between the top features selected by SHAP and those identified by LIME. Compared with SHAP approach, feature importance identified by LIME is more straightforward and intuitive.

*Counterfactual explanations explaining why not a different prediction was made*

We answered the "Why not" question using counterfactual explanations, which identifying the minimum changes necessary to obtain a different outcome. We restricted the variation on preoperative laboratory features only and limited their variation within a specified range, from the 1th percentile and 99$^{th}$ percentile values of training cohort. Figure 3 and eFigure 4 presented the counterfactual explanations for two example patients. For the patient at high risk (Figure 3), originally abnormal hemoglobin, serum calcium, serum calcium, serum potassium, and white blood cell count were required to return to normal values to reduce the risk from 74% to 49%. Conversely, for the patient with low risk (eFigure 4), in order to flip the outcomes, altering several originally normal laboratory measurements to abnormal values was required to reverse the outcomes.

*Model cards documenting how to use the model*

We developed a model card to disclose the operational and methodological aspects information of our MySurgeryRisk model (Figure 4). It describes the data source, the context in which models are intended to be used, references, distribution of demographic variables and outcomes, as well as performance.

We augmented model cards with global feature importance, which measures the overall impact of each feature on the model's predictions (Figure 5). Primary procedure code was identified to have the most contribution in general. Additional key features included attending surgeon, preoperative serum calcium, surgery type and preoperative serum glucose, which ranked among the top five important features. We also conducted analysis based on sex (female vs male), race (African American vs non-African American) and age (age⩽65 vs >65 years old) groups. We observed similar feature importance among sex group and race group, but slightly different feature importance for age group.

*Interactive manipulation of feature values explaining "What if" question*

We developed an interactive user interface that enables users to modify feature values. After users update the values, we run inference to provide updated prediction risks. Figure 6 displays the user interface for an example patient, showing the updated risk score and the altered feature values at the top once the user submits the updated data. For example, the example patient's predicted risk score has decreased from 12.2% to 11.7% after lowering the body mass index and adjusting for comorbidities such as diabetes and hypertension.

*Similar patients answering "What else" question*

We provided model's predictions on similar instances. We identified similar patients by matching principal features. That is, two patients were considered similar if their age difference was five years or less, had same race and gender characteristics, scheduled for the same type of surgery and exhibited at least 60% similarity in comorbidities. Figure 7 present the prediction results on similar patients for an example patient. It shows both the current patient's predicted risks and the average predicted risks for similar patients.

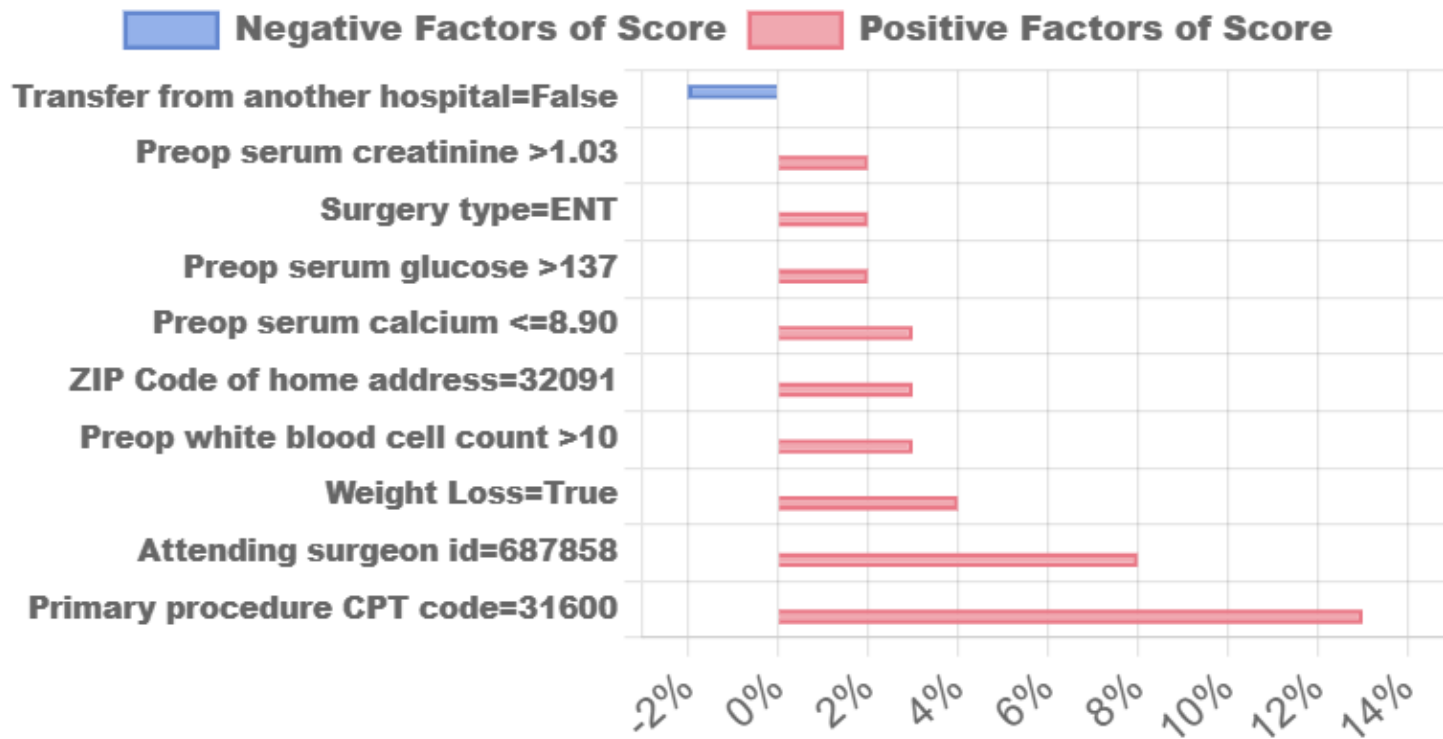

Figure 2. Feature importance provided by Local Interpretable Model-agnostic Explanations (LIME) approach explain why the patient has high risk of developing prolonged mechanical ventilation.

## Suggestions from Model

If the following features were changed, the risk of developing prolonged mechanical ventilation would decrease from **74%** to **49%**

| Features | Raw Value | New Value |
|---|---|---|
| Hemoglobin | 7.1 g/dL | 12.4 g/dL |
| Serum Calcium | 7.7 mg/dL | 8.9 mg/dL |
| Serum Glucose | 319 mg/dL | 119 mg/dL |
| Serum Potassium | 3.2 mmol/L | 4.2 mmol/L |
| White blood cell count | 10.1 × 10^9/L | 8.4 x × 10^9/L |

**Figure 3. Counterfactual explanation explains why not the patient was at low risk of developing prolonged mechanical ventilation.**

## Overview
MySurgeryRisk predicts the risk for eight main postoperative complications (acute kidney injury, sepsis, venous thromboembolism, prolonged intensive care unit stay > 48 hours, prolonged mechanical ventilation > 48 hours, wound, neurologic, and cardiovascular complications).

## Source of Data & Patients
Using the University of Florida Health (UFH) Integrated Data Repository as Honest Broker for data deidentification we have created a single-center perioperative longitudinal cohort that integrated electronic health record (EHR) data. We included all inpatient operative procedures performed between June 1, 2014 and September 20, 2020.

## References
- https://www.ncbi.nlm.nih.gov/pmc/articles/PMC6110979/
- https://jamanetwork.com/journals/jamanetworkopen/article-abstract/2792367

## Intended Users
- Medical professionals
- Machine learning researchers

## Use Cases
- Input: electronic health records data including demographics, admission information, scheduled surgery information, comorbidity, laboratory measurements and medications within one year prior to surgery
- Output: Risk of eight postoperative complications and death (value range [0, 1])
- Machine learning model: Random Forest
- Usage: Before surgery, MySurgeryRisk will estimate the risk of complications and inform the doctor.

### Training Data Cohort

| Features | Development Cohort | Validation Cohort |
|---|---|---|
| Number of Patients, N | 41,812 | 19,132 |
| Number of Encounters, N | 52,117 | 22,300 |
| **Demographic Information** | | |
| Age, Years, Mean (SD) | 56 (17) | 58 (17) |
| **Sex, N (%)** | | |
| Male, N (%) | 26,046 (50) | 10,927 (49) |
| Female, N (%) | 26,071 (50) | 11,373 (51) |

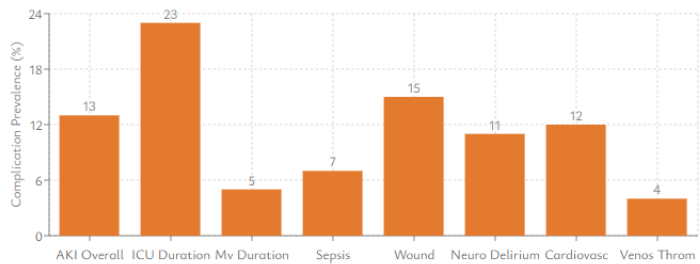

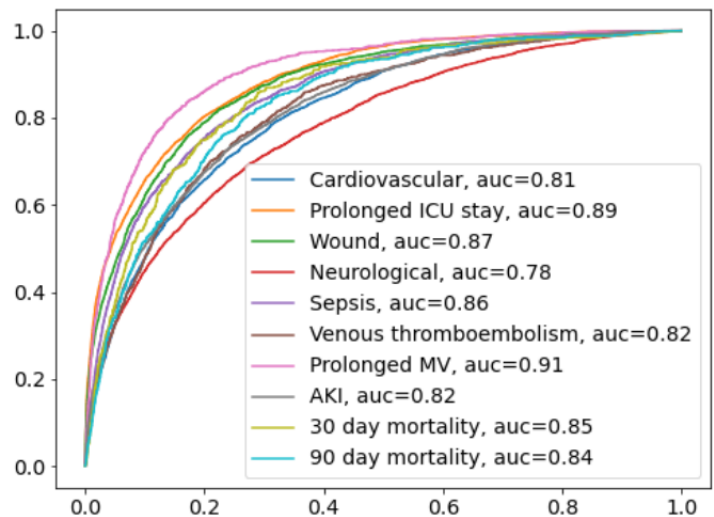

Figure 4. Model cards document the operational and methodological aspects information of our MySurgeryRisk model.

| Overall | AKI | ICU/Admission | **Mechanical Ventilation** | Sepsis | Wound | Neuro and Delrium | Cardiovascular | Venous Thromboembolism |

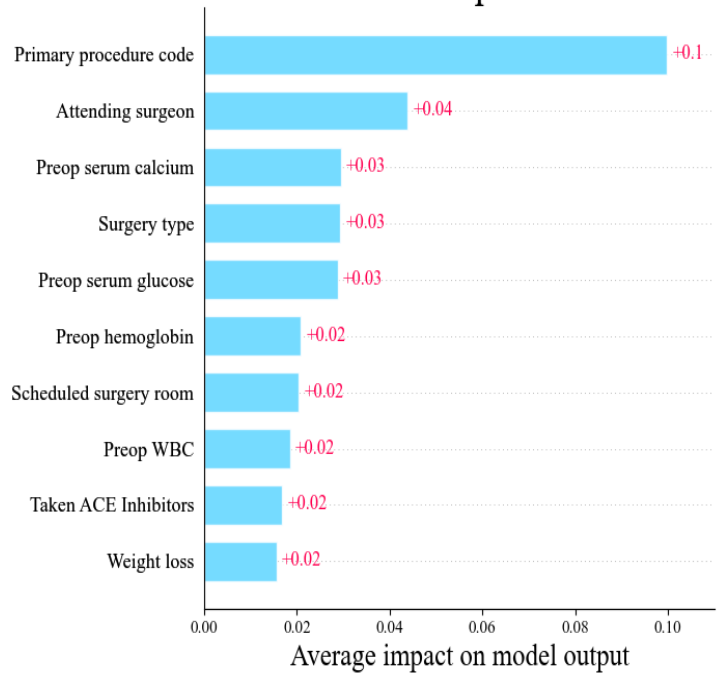
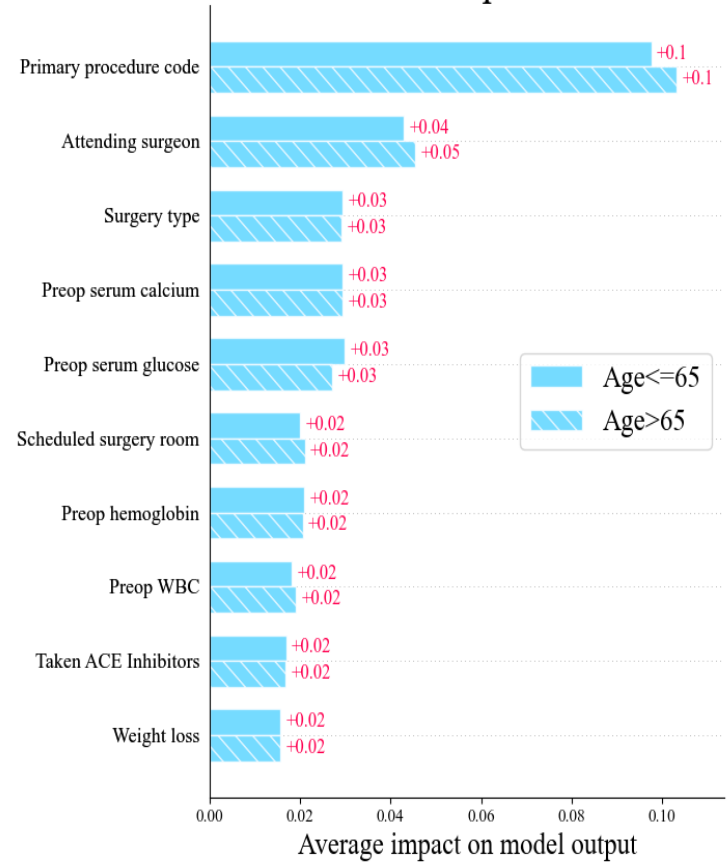

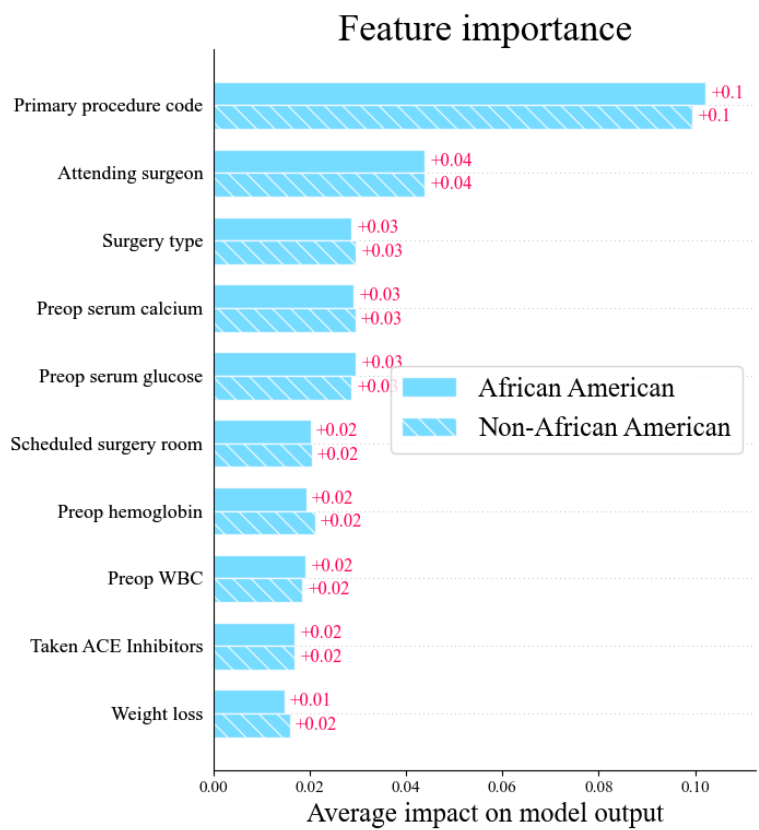 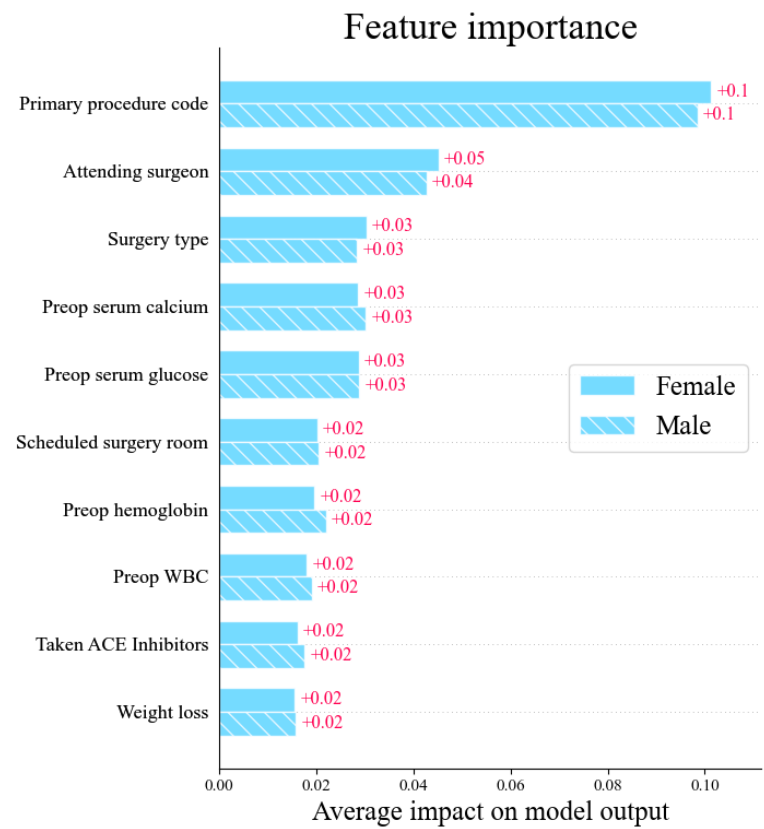

**Figure 5.** Global feature importance for all patients, age<=65 and age >65 patient groups, African American and Non-African American patient groups, and female and male patient groups.

Figure 6. Interactive feature manipulation interface for an example patient. (A) Prediction results of original feature values. (B) Prediction results of modified feature values.

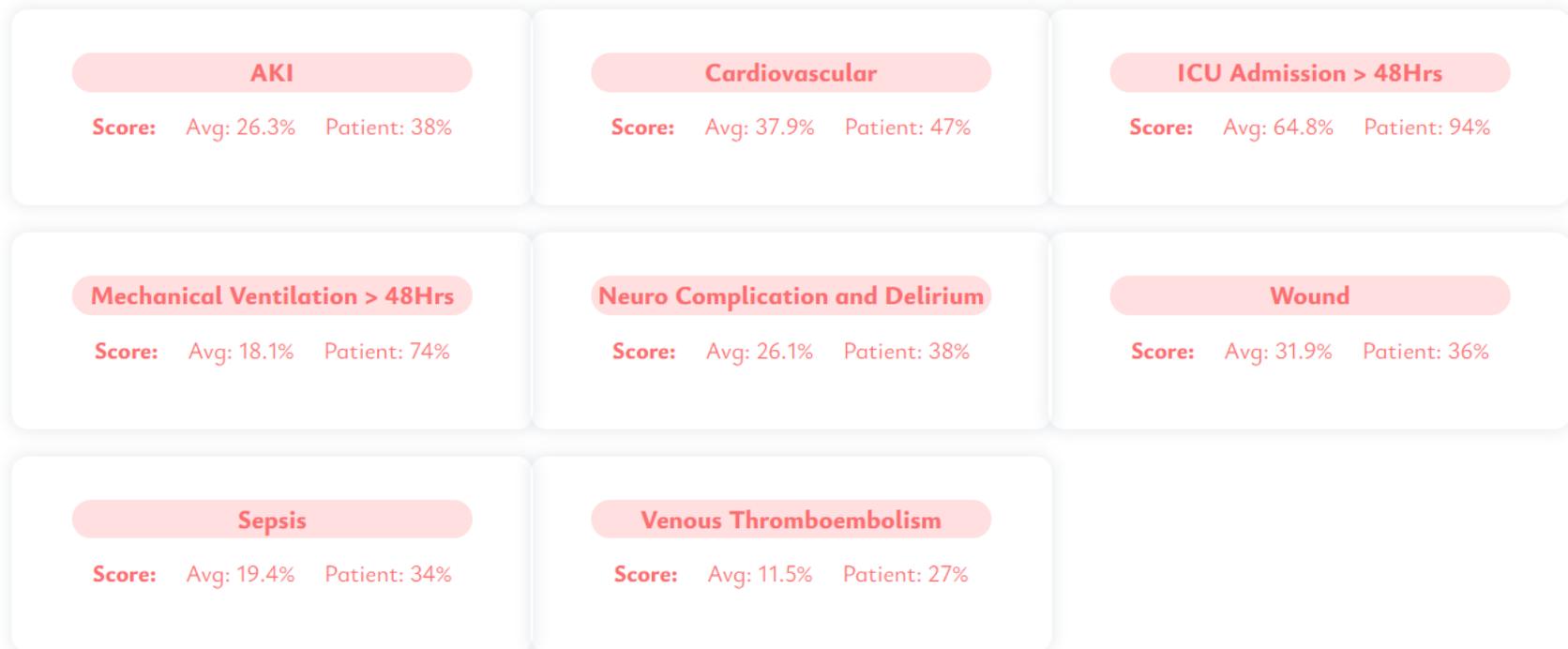

Figure 7. Risk predictions of developing prolonged mechanical ventilation on similar patients for an example patient.

**Discussion**

We have proposed an XAI framework aiming at enhancing the explainability and transparency of AI model. This framework addresses five critical questions: "why?" a particular decision was made, "why not?" a different outcome was not chosen, "how?" the model functions and how to use it, "what if?" feature values are altered and the resulting impact on predictions, and 'what else?' about other similar patients. These questions are systematically answered within the scope of our XAI framework by incorporating approaches such as LIME, SHAP, counterfactual explanations, model cards and interactive interface.

Local feature importance measures how features affect the prediction for a specific sample, explaining why a model made a particular decision for a single data point. This importance can vary significantly across samples and is crucial for understanding individual predictions in high-stakes scenarios such as medical diagnoses and loan approvals. For example, in a study at Mount Sinai Hospital used convolutional neural networks (CNNs) to identify high-risk patients based on their X-ray images. The model was successful when used internally, but when applied to external institutions, its performance significantly decreased. Further investigation showed that the model was basing its prediction on hardware meta-data linked to the x-ray machine exclusively used to image high-risk patients at Mount Sinai instead of accurately identifying high-risk patients.[11, 12] Global feature importance measures the importance of features across the entire dataset or model, on the other hand, quantifies the significance of features across the entire dataset or model, providing a broad overview of which features are most influential for the model's predictions overall. It is more useful in feature engineering and model debugging. In our MySurgeryRisk model, primary procedure code was identified as the most important feature, consistent with our previous finding.[4] Counterfactual explanations provide solutions to achieve a different outcome. For patients identified as high risk, counterfactual explanations may suggest specific changes in health behaviors, treatment plans, or environmental factors that could theoretically lower their risk. For instance, for a patient with a high risk of requiring prolonged MV, a counterfactual explanation might illustrate how controlling the weight loss could shift their risk status to lower levels. For patients identified as low risk, counterfactual explanations may assist healthcare providers in better monitoring and managing those factors to prevent the patient's risk from increasing. Our interactive feature manipulation interface complements these counterfactual explanations by providing a more engaging and flexible user experience, although it's important to note that counterfactual explanations are not guaranteed for every sample. Comparing a patient to a cohort of similar individuals, healthcare providers can predict outcomes more accurately and customize treatments that are more likely to be effective for that specific patient. Existing studies often limit to providing global and local feature importance. Our XAI framework providing comprehensive explainability knowledge represents the initial steps towards transparent AI and clinical adoption.

The study had several limitations. First, the model utilized random forest algorithm, other gradient based approaches, such as Integrated Gradients and Attention Maps, were not appropriate. This limitation restricted the scope of exploration and generalizability. Second, the proposed XAI framework and designed prototype were not clinically validated, raising concerns about their practical applicability in real-world medical settings.

## **Conclusions**

In this study, we proposed an XAI framework designed to answer five critical questions "*why, why not, how, what if, and what else*", aiming at enhancing the explainability and transparency of AI model. We showcased an XAI interface prototype that adheres to this framework for predicting major postoperative complications. This initial implementation yielded valuable insights into the extensive explanatory potential our XAI framework can provide.

Moving forward, our future work will focus on a thorough XAI evaluation, refinement of our interface, and the development of a real-time XAI MySurgeryRisk system.

**eFigure 1. Feature importance provided by Local Interpretable Model-agnostic Explanations (LIME) approach explain why the patient has high risk of developing prolonged mechanical ventilation.**

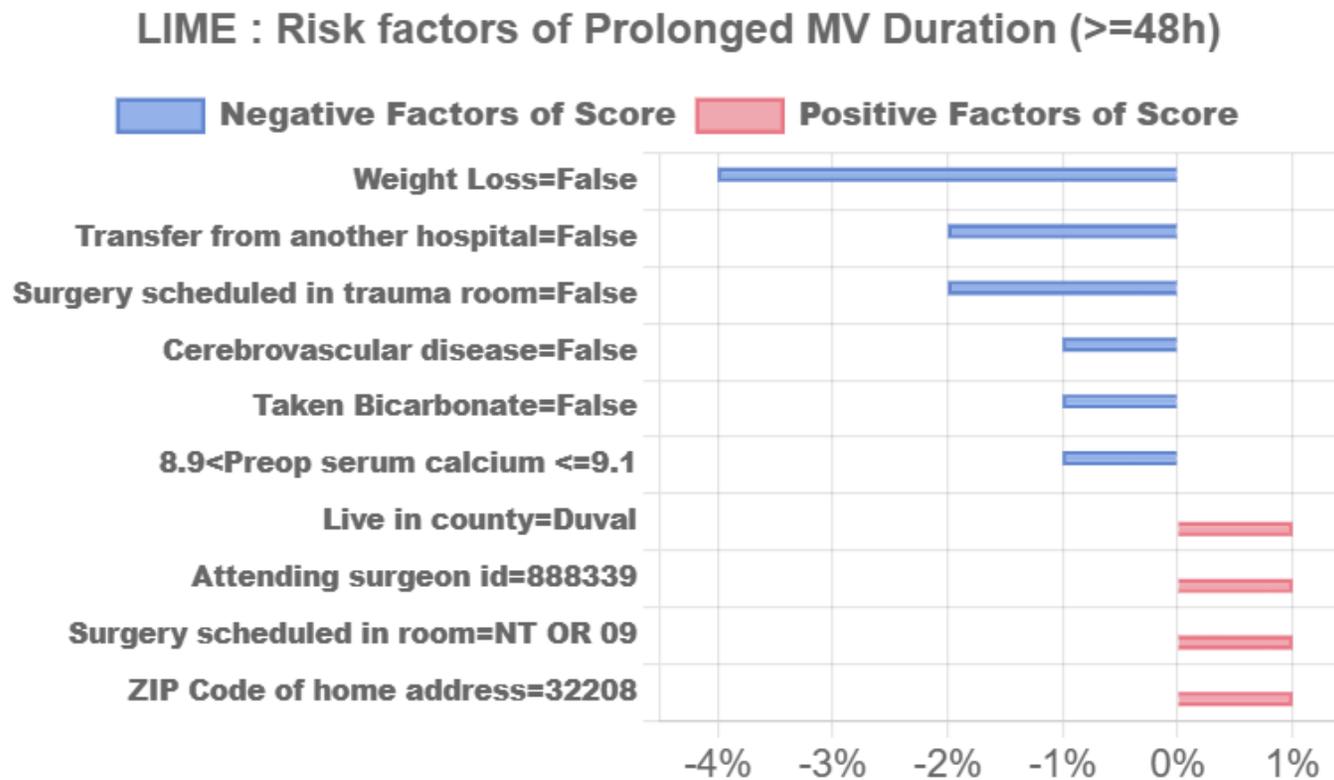

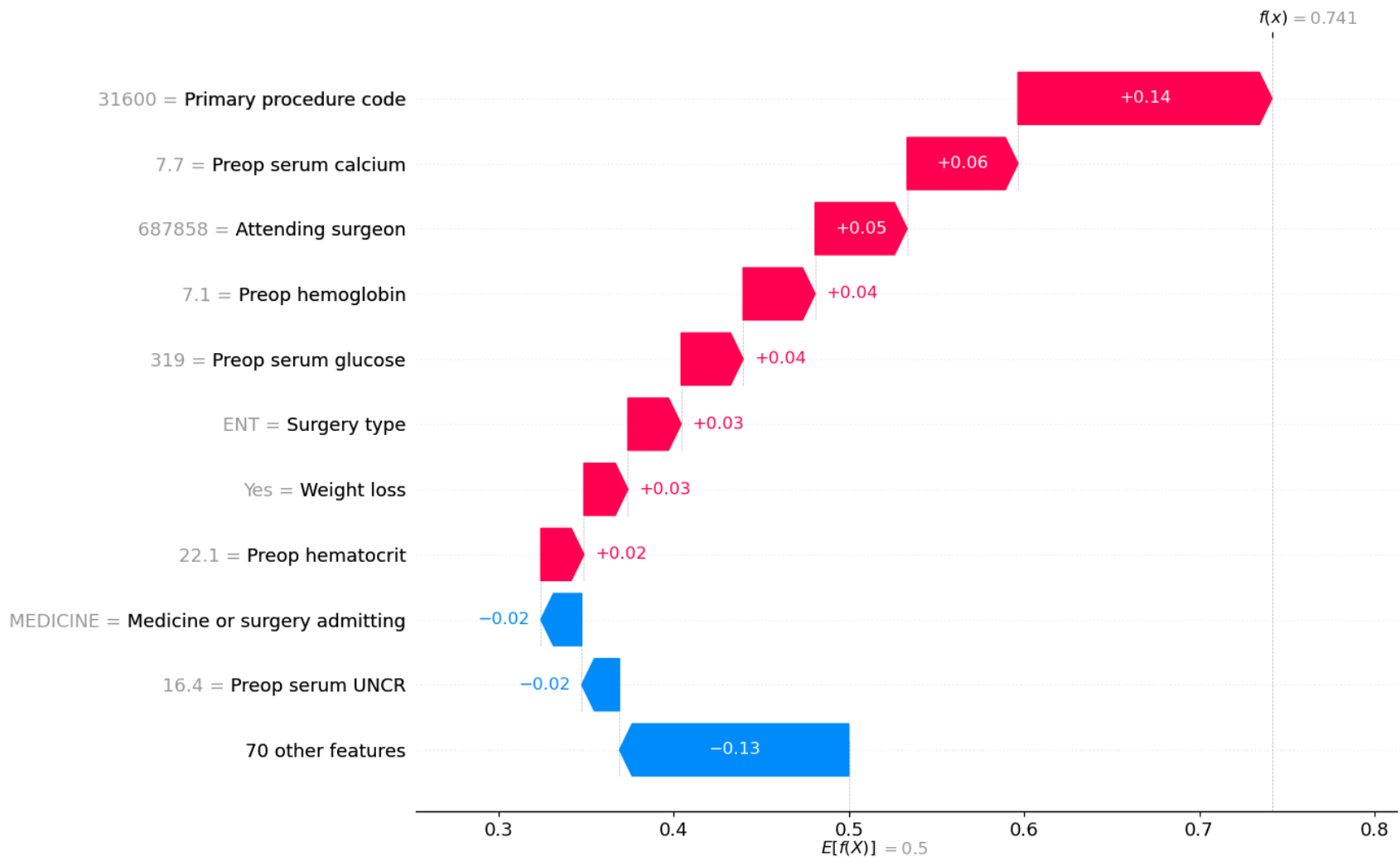

eFigure 2. Feature importance provided by SHapley Additive exPlanations (SHAP) approach explain why the patient has high risk of developing prolonged mechanical ventilation.

eFigure 3. Feature importance provided by SHapley Additive exPlanations (SHAP) approach explain why the patient has low risk of developing prolonged mechanical ventilation.

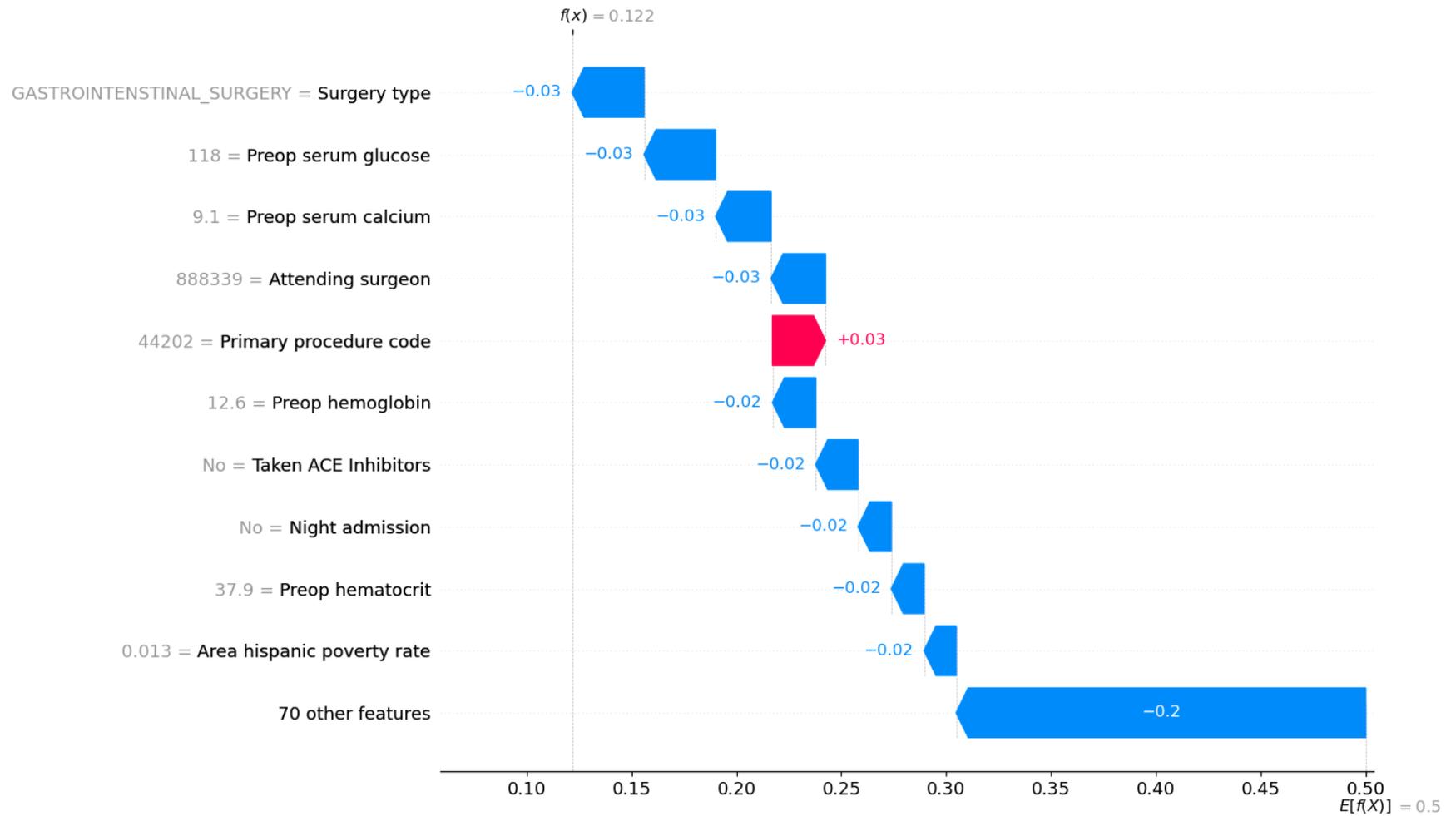

**eFigure 4. Counterfactual explanation explains why not the patient was at high risk of developing prolonged mechanical ventilation.**

## Suggestions from Model

If the following features were changed, the risk of developing prolonged mechanical ventilation would increase from **12%** to **50%**

| Features | Raw Value | New Value |
|---|---|---|
| Blood Urea Nitrogen | 16 mg/dL | 74 mg/dL |
| Hemoglobin | 12.6 g/dL | 6.3 g/dL |
| Serum Calcium | 9.1 mg/dL | 7.1 mg/dL |
| Serum Glucose | 118 mg/dL | 334.5 mg/dL |
| BUN-to-Creatinine Ratio | 17.2 | 47.3 |

**eTable 1. Input features used in models.**

| Feature | Model published | Model in this study | type |
|---|---|---|---|
| Sex | X | X | Binary |
| Native Language Spoken | X | X | Binary |
| Scheduled post operation location | X | | Binary |
| Scheduled room is trauma room or not | X | X | Binary |
| rural or city at patient residential area | X | X | Binary |
| emergent or elective admission | X | X | Binary |
| Anesthesia Type | X | X | Binary |
| Yes, if the admission we happened at night | X | X | Binary |
| CKD status at admission | X | X | Categorical (Binary for AKI) |
| Myocardial Infarction | X | X | Binary |
| Congestive Heart Failure | X | X | Binary |
| Cerebrovascular Disease | X | X | Binary |
| Chronic Pulmonary Disease | X | X | Binary |
| Peripheral Vascular Disease | X | X | Binary |
| Cancer | X | X | Binary |
| Liver Disease | X | X | Binary |
| Valvular Disease | X | X | Binary |
| Coagulopathy | X | X | Binary |
| Weight Loss | X | X | Binary |
| alcohol abuse or drug | X | X | Binary |
| Fluid and electrolyte disorders | X | X | Binary |
| Chronic anemia | X | X | Binary |
| Hypertension | X | X | Binary |
| Obesity | X | X | Binary |
| Diabetes | X | X | Binary |
| Metastatic Carcinoma | X | X | Binary |
| Depression | X | X | Binary |
| Indicator of receiving Betablockers within one year before admission date | X | X | Binary |
| Indicator of receiving Diuretics within one year before admission date | X | X | Binary |
| Indicator of receiving statin within one year before admission date | X | X | Binary |

| Feature | Model published | Model in this study | type |
|---|---|---|---|
| Indicator of receiving Aspirin within one year before admission date | X | X | Binary |
| Indicator of receiving ACE Inhibitors within one year before admission date | X | X | Binary |
| Indicator of receiving vasopressors or inotropes within one year before admission date | X | X | Binary |
| Indicator of receiving Bicarbonate within one year before admission date | X | X | Binary |
| Indicator of receiving Antiemetic within one year before admission date | X | X | Binary |
| Indicator of receiving Aminoglycosides within one year before admission date | X | X | Binary |
| Indicator of receiving Vancomycin within one year before admission date | X | X | Binary |
| No of nephrotoxic drugs received within one year before admission date | X | X | Binary |
| Race | X | X | Categorical |
| Ethnicity | X | X | Categorical |
| Insurance paying the bills | X | X | Categorical |
| Admission Source | X | X | Binary |
| Scheduled surgery room | X | X | Categorical |
| Smoking Status | X | X | Categorical |
| Automated urinalysis, urine protein presence within 365 days prior to surgery, mg/dL | X | | Categorical |
| Automated urinalysis, urine glucose within 7 days prior to surgery, mg/dL | | X | Categorical |
| Automated urinalysis, urine hemoglobin within 7 days prior to surgery, mg/dL | X | X | Categorical |
| Automated urinalysis, urine hemoglobin within 8-365 days prior to surgery, mg/dL | X | | Categorical |
| Number of urine hemoglobin tests within 8-365 days prior to surgery | X | | Categorical |
| Surgery Type | X | X | Categorical |

| Feature | Model published | Model in this study | type |
|---|---|---|---|
| Marital Status | X | X | Categorical |
| ZIP Code of home address | X | X | Categorical |
| County of home address | X | X | Categorical |
| Admission Day | X | X | Categorical |
| Admission Month | X | X | Categorical |
| ID of Attending Surgeon |  | X |  |
| medicine or surgery admitting | X | X | Categorical |
| Charlson comorbidity index | X | X | Categorical |
| Current Procedural Terminology code of the primary procedure | X | X | Categorical |
| Age | X | X | Numerical |
| Body Mass Index | X | X | Numerical |
| Distance of residence to hospital, km | X | X | Numerical |
| total population at patient residential area |  | X |  |
| median total income at patient residential area, USD | X | X | Numerical |
| prevalence of African American residents living below poverty at patient residential area, % | X | X | Numerical |
| prevalence of Hispanic residents living below poverty at patient residential area, % | X | X | Numerical |
| Prevalence of residents living below poverty at patient residential area, % | X | X | Numerical |
| Time from Admission to Surgery, days | X |  | Numerical |
| Reference estimated glomerular filtration rate | X | X | Numerical |
| Min hemoglobin within 7 days prior to surgery, g/dl | X | X | Numerical |
| Max hemoglobin within 7 days prior to surgery, g/dl | X |  | Numerical |
| Average of hemoglobin within 7 days prior to surgery, g/dl | X |  | Numerical |
| Variance of hemoglobin within 7 days prior to surgery, g/dl | X |  | Numerical |

| Feature | Model published | Model in this study | type |
|---|---|---|---|
| Number of hemoglobin tests within 7 days prior to surgery | X | | Numerical |
| Min hemoglobin within 8-365 days prior to surgery, g/dl | X | | Numerical |
| Max hemoglobin within 8-365 days prior to surgery, g/dl | X | | Numerical |
| Average of hemoglobin within 8-365 days prior to surgery, g/dl | X | | Numerical |
| Number of hemoglobin tests within 8-365 days prior to surgery | X | | Numerical |
| Min of Serum Calcium, mmol/L | X | X | Numerical |
| Max of Serum Calcium, mmol/L | X | | Numerical |
| Average of Serum Calcium, mmol/L | X | | Numerical |
| Variance of Serum Calcium, mmol/L | X | | Numerical |
| Count of Serum Calcium test | X | | Numerical |
| Max of anion gap in blood, mmol/L | | X | Numerical |
| Average of anion gap in blood, mmol/L | X | | Numerical |
| Count of anion gap in blood test | X | | Numerical |
| Min of White Blood Cell in blood, thou/uL | X | | Numerical |
| Max of White Blood Cell in blood, thou/uL | X | X | Numerical |
| Average of White Blood Cell in blood, thou/uL | X | | Numerical |
| Variance of White Blood Cell in blood, thou/uL | X | | Numerical |
| Count of White Blood Cell in blood test | X | | Numerical |
| Min of Hematocrit in blood, % | X | X | Numerical |
| Average of Hematocrit in blood, % | X | | Numerical |
| Variance of Hematocrit in blood, % | X | | Numerical |
| Count of Hematocrit in blood test | X | | Numerical |
| Max of Serum Red Blood Cell, Million/uL | X | X | Numerical |
| Average of Serum Red Blood Cell, Million/uL | X | | Numerical |
| Max of the amount of hemoglobin relative to the size of the cell in blood, g/dL | X | X | Numerical |

| Feature | Model published | Model in this study | type |
|---|---|---|---|
| Average of the amount of hemoglobin relative to the size of the cell in blood, g/dL | X | | Numerical |
| Min of Glucose in blood, mg/dL | X | | Numerical |
| Max of Glucose in blood, mg/dL | X | X | Numerical |
| Average of Glucose in blood, mg/dL | X | | Numerical |
| Count of Glucose in blood test | X | | Numerical |
| Min of Serum CO2 , mmol/L | X | X | Numerical |
| Max of Serum CO2, mmol/L | X | | Numerical |
| Average of Serum CO2, mmol/L | X | | Numerical |
| Variance of Serum CO2, mmol/L | X | | Numerical |
| Count of Serum CO2 test | X | | Numerical |
| Min of Urea nitrogen in blood, mg/dL | X | | Numerical |
| Max of Urea nitrogen in blood, mg/dL | X | X | Numerical |
| Average of Urea nitrogen in blood, mg/dL | X | | Numerical |
| Variance of Urea nitrogen in blood, mg/dL | X | | Numerical |
| Count of Urea nitrogen in blood test | X | | Numerical |
| Min of Urea Nitrogen-Creatinine ratio | X | | Numerical |
| Max of Urea Nitrogen-Creatinine ratio | X | X | Numerical |
| Average of Urea Nitrogen-Creatinine ratio | X | | Numerical |
| Variance of Urea Nitrogen-Creatinine ratio | X | | Numerical |
| Count of Urea Nitrogen-Creatinine ratio | X | | Numerical |
| Max of Serum Sodium, mmol/L | X | X | Numerical |
| Average of Serum Sodium, mmol/L | X | | Numerical |
| Count of Serum Sodium test | X | | Numerical |
| Average of Potassium in serum, mmol/L | X | | Numerical |
| Min of Potassium in serum, mmol/L | | X | Numerical |
| Count of Potassium in serum test | X | | Numerical |

| Feature | Model published | Model in this study | type |
|---|---|---|---|
| Max of Red cell distribution width in Blood, % | X | X | Numerical |
| Average of Red cell distribution width in Blood, % | X | | Numerical |
| Variance of Red cell distribution width in Blood, % | X | | Numerical |
| Min of platelet in blood, thou/uL | X | X | Numerical |
| Max of platelet in blood, thou/uL | X | | Numerical |
| Average of platelet in blood, thou/uL | X | | Numerical |
| Variance of platelet in blood, thou/uL | X | | Numerical |
| Min of Serum creatinine, mg/dL | X | | Numerical |
| Max of Serum creatinine, mg/dL | X | X | Numerical |
| Average of Serum creatinine, mg/dL | X | | Numerical |
| Variance of Serum creatinine, mg/dL | X | | Numerical |
| Count of Serum creatinine test | X | | Numerical |
| Max of chloride in Serum, mmol/L | X | | Numerical |
| Average of chloride in Serum, mmol/L | X | | Numerical |
| Variance of chloride in Serum, mmol/L | X | | Numerical |
| Count of chloride in Serum test | X | | Numerical |